# Ballistic Graphene Nanoribbon MOSFETs: a full quantum real-space simualtion study

Gengchiau Liang[*],

Electrical and Computer Engineering, National University of Singapore, Singapore 117576

Neophytos Neophytou, Mark S. Lundstrom, and

School of Electrical and Computer Engineering, Purdue University, West Lafayette, Indiana 47907-1285

Dmitri E. Nikonov

Technology and Manufacturing Group, Intel Corp., SC1-05, Santa Clara, California 95052

---

[*] Email address: elelg@nus.edu.sg

**Abstract**

A real-space quantum transport simulator for carbon nanoribbon (CNR) MOSFETs has been developed. Using this simulator, the performance of carbon nanoribbon (CNR) MOSFETs is examined in the ballistic limit. The impact of quantum effects on device performance of CNR MOSFETs is also studied. We found that 2D semi-infinite graphene contacts provide metal-induced-gap-states (MIGS) in the CNR channel. These states would provide quantum tunneling in the short channel device and cause Fermi level pining. These effects cause device performance degradation both on the ON-state and the OFF-state. Pure 1D devices (infinite contacts), however, show no MIGS. Quantum tunneling effects are still playing an important role in the device characteristics. Conduction due to band-to-band tunneling is accurately captured in our simulations. It is important in these devices, and found to dominate the off-state current. Based on our simulations, both a 1.4nm wide and a 1.8nm wide CNR with channel length of 12.5nm can outperform ultra scaled Si devices in terms of drive current capabilities and electrostatic control. Although subthreshold slopes in the forward-bias conduction are better than in Si transistors, tunneling currents are important and prevent the achievement of the theoretical limit of 60mV/dec.

# I. INTRODUCTION

Graphite-related materials such as fullerenes, graphene, and carbon nanotubes have generated considerable interest due to unique electronic and optoelectronic properties. For instance, it has been demonstrated that carbon nanotubes (CNTs) exhibited quasi-ballistic conduction and can function as transistors[1], light-emitters[2,3], and sensors[4]. Similarly, 2D graphene (i.e. monolayer graphite) sheets have been shown to possess very high carrier mobility. Unlike CNTs, the metallic nature of this material with zero bandgap prohibits the use of 2D graphene sheets as semiconductor devices. Recently graphene sheets have been patterned into narrow nanoribbons[5,6]. Due to quantum confinement, graphene nanoribbons can have bandgap, depending on their width and orientation relative to the graphene crystal structure[7,8]. Similar to CNTs, both semiconducting and metallic properties can be achieved by armchair CNRs and zigzag CNRs[7,8], respectively. FET type devices based on the armchair CNRs have been studied both experimentally[9] and theoretically[8,10,11,12]. Recent studies using a semi-classical transport model predicted that CNR MOSFETs might outperform traditional Si MOSFETs[8] and could have a competitive ON-current performance with CNT MOSFETs[11,12]. However, little is known about how quantum effects affect the device performance of the CNR MOSFETs. Among several quantum effects that affect this type of structures, tunneling plays an important role since it would degrade the device performance by reducing the ratio of $I_{ON}$ to $I_{OFF}$, and by increasing the subthreshold slope.

In this paper, we present a full real-space quantum transport simulator based on the Non-equilibrium Green's Function (NEGF) approach, self-consistently coupled to a 3D Poisson's equation solution for treatment of the electrostatics. We study the contact effect of armchair CNR devices by considering both 1D perfect contacts (of width equal to that of the channel), and 2D semi-infinite contacts. We found that due to the metal-induced gap states (MIGS) from 2D contacts, localized states appear in the middle of the bandgap. These levels will pin the Fermi level and contribute to tunneling when the channel is short, therefore degrading the device performance. The armchair CNR MOSFETs with 1D perfect contacts show a clear bandgap with no MIGS and is the focus of this study.

We consider the 1.4 nm wide and 1.8 nm wide CNR MOSFETs with a channel length of 12.5 nm to investigate the performance of an armchair CNR MOSFETs. The subthreshold swing (SS) of these two CNR MOSFETs is around 68mV/decade and 74mV/decade respectively. The drain-induced-barrier-lowering (DIBL) is 30mV/V and 60mV/V respectively, which is smaller than the theoretically expected value of a double gate 10nm-scale Si MOSFET [13]. The reason of performance enhancement in CNR MOSFET can be attributed to nature of the CNR device, consisting of a single monolayer of carbon atoms, (an ultimate ultra-thin-body, double gate MOSFET). The electrostatic control of such device can be superior to other planar transistors, since it eliminates degrading electrostatic effects associated with 2D geometries. The subthreshold swing of 60 mV/decade, however, cannot be achieved in this operation mode. The light effective mass of CNRs, in combination with their small bandgap and the short channel length, enhance carrier tunneling through the barrier from the source to the drain. This effect degrades the OFF-state device performance. On the other hand, this tunneling process can be utilized to other operating schemes, such as in band-to-band-tunneling (BTBT) devices. The subthreshold swing of the band-to-band-tunneling CNT MOSFET has been demonstrated to be smaller than the fundamental thermal limit $k_B T/q$, (60 mV/decade at room temperature)[14,15]. Based on our simulations, the CNR MOSFETs also show the potential to be used in the BTBT regime.

## II. APPROACH

The device characteristics of CNR MOSFETs are simulated using 3D Poisson's equation coupling a quantum transport model based on Non-Equilibrium Greens' Function (NEGF) formalism. The Poisson equation is solved in three-dimensional (3D) coordinates using the method of moments (boundary element method)[16]. In the quantum transport model, the bandstructure of the device is described by a Hamiltonian $H_D$ using simple π-orbital tight-binding model[17] between nearest neighbor. The self-consistent potential $\phi$ coming from Poisson's equation accounts for electron-electron interactions. In the next section we describe these two parts of the simulation process in detail.

*Quantum transport simulation:* The Hamiltonian of the device is taken in the approximation of π orbital tight binding (TB) in the form

$$H_D = \sum_{ij} t |i\rangle \langle j|, \tag{1}$$

where summation is performed over nearest neighbor atom pairs, with zeros for the on-site (diagonal) elements [15]. The coupling parameter is taken to be $t=3$ eV[17]. The Green's function for the device is determined by

$$G(E) = \left[ \left( E + i0^+ \right) I - H - \varSigma_S - \varSigma_D \right]^{-1}, \tag{2}$$

where $I$ is the identity matrix, $\varSigma_S$ and $\varSigma_D$ are the self energies for the left (source) and right (drain) reservoirs. The local density of states resulting from source (drain) injected states is calculated using

$$D_{S(D)}(E) = (1/2\pi)G\Gamma_{S(D)}G^+, \tag{3}$$

where $\Gamma_{S(D)} = i\left(\Sigma_{S(D)} - \Sigma_{S(D)}^+\right)$, is broadening due to the source/drain contacts. The electron correlation function is computed according to

$$G^n(E) = G\Sigma^{in}G^+, \tag{4}$$

where, for ballistic conduction, the in-scattering function corresponding to the contacts $\Sigma_{S(D)}^{in}(E) = \Gamma_{S(D)}(E)f(E, E_{FS(D)})$ is determined by the corresponding fermi distribution function. Charge density is calculated by integrating the electron correlation function $G^n(E)$ over energy as follows [18,19,20].

$$\rho_j = e\int_{-\infty}^{\infty}\frac{dE}{2\pi}G_j^{\ n}(E) \tag{5}$$

After convergence, the current is calculated by[18,19,20]

$$I_{j \to j+1} = \frac{ie}{\hbar}\int_{-\infty}^{\infty}\frac{dE}{2\pi}[H_{j,j+1}G_{j+1,j}^n(E) - HG_{j,j+1}^n(E)] \ . \tag{6}$$

The self-energies, $\Sigma_S$ and $\Sigma_D$, for the left and right reservoirs are designed to model infinite length reservoirs using the Sancho-Rubio iterative method[21]. We applied the NEGF approach with the Sancho-Rubio iterative method to calculate the density of states, $DOS(E)$, and transmission, $T(E)$, for an infinite graphene sheet. The inset of Fig. 1 shows a schematic of the partition of the device into the channel (an elementary lattice cell consisting of two atoms, in this case) and the contacts in the simulations. It also shows the simulated $DOS(E)$ and $T(E)$ of the infinite 2D graphene sheet. It has excellent agreement with previous studies[17], and demonstrates that the computed self-energy gives the accurate description of an infinitely extended reservoir. In this work, we implement two options for boundary conditions depending on the nature of the contacts that we want to employ. Specifically, 1D contacts are implemented when we assume that the nanoribbon extends to infinity having the same width as in the channel. In the 2D case, we assume that the contact is a 2D graphene half-plane sheet that is connected to the channel at the left and the right contacts, as indicated in the schematics of Fig. 2. There are two signature differences between 1D contact and 2D contact CNR devices. First, due to the interface mismatch, strong oscillations appear in T(E) and DOS(E). The effects will be reduces if the channel length (*L*) is much larger than its width (*W*). Moreover, for the 2D contacts, a localized state in the middle of bandgap appears due to the metallic contact behavior of the graphene sheet (metal induced gap states- MIGS). When the channel is long enough, these states will only be localized appear at the interface and do not contribute to current. However, they can still cause Fermi level pinning and degrade device performance.

*3D Poisson equation*: We employ a 3D electrostatics solution for the device using the method of moments (boundary element method)[15,23]. Three-dimensional treatment is necessary in our case because there is no obvious symmetry in the structure that can reduce the treatment of the device to a lower dimensionality with reasonable approximations. The main difference between the nanoribbon devices and conventional Si MOSFETs is that in the Si the width of the channel alters the transport properties of the MOSFET in a trivial way (i.e. if the width is doubled, the current doubles). In the case of the ribbons, however, this is not the case since the electronic properties of the ribbon are a sensitive function of its width. The method of moments is a suitable method for treatment of the electrostatics of these type of devices.

The charge and the potential are separated into the graphene-device *(n_D, Φ_D)* and the gate/source/drain-boundary *(n_B, Φ_B)* parts. Grid points are placed only on the regions of the structure on which charge can reside. For the boundary, these are only the surfaces of the source, drain and gate electrodes. Device grid points are placed on all the graphene atoms, on which charge can reside. The elements on the boundary surfaces are assumed to be rectangles with differential element $\Delta S$ and on the device (graphene sheet) circles with radius $R_{eff}$. The charge distribution is assumed to be composed of point charges in the center of the differential element. The potential at the boundary and device is related to the charge density in the structure by:

$$\begin{Bmatrix} \Phi_D \\ \Phi_B \end{Bmatrix} = K(r; r') \begin{Bmatrix} n_D \\ n_B \end{Bmatrix} = \begin{bmatrix} A & B \\ C & D \end{bmatrix} \begin{Bmatrix} n_D \\ n_B \end{Bmatrix}, \tag{7}$$

where K is the electrostatic Kernel of the device geometry under examination, A, B are the matrices that describe the contribution of device and boundary charge respectively on the potential on the device, and C, D describe the contribution of device and boundary charge respectively on the potential on the boundary. Under the point charge approximation, the potential $\Phi_i$ at each element consists of an on-site potential $\Phi_{ii}$ and a summation term for the contribution to that potential of all the charges in the system, as follows

$$\Phi_i = \Phi_{ii} + \sum_{j \neq i}^{N} \frac{\rho_{sj} \Delta S_j}{4\pi\varepsilon |r_i - r_j|}, \tag{8}$$

$$\Phi_{ii} = \frac{Q_{tot}}{4\pi\varepsilon R_{eff}} = \frac{\rho_{si} \Delta S_i}{4\pi\varepsilon \sqrt{\Delta S_i / 4}} = \frac{\rho_{si} \sqrt{\Delta S_i}}{2\pi\varepsilon}, \tag{9}$$

where $\Phi_{ii}$ is the on-site energy of the rectangular surfaces on the boundary nodes. The effect of charge imaging due to the presence of different dielectric regions in the device is also taken into account. Once the electrostatic Kernel $K(r, r')$ is build, the potential on the device can be calculated from (8) to be:

$$\Phi_D = (A - BD^{-1}C) \, n_D \; + \; BD^{-1} \, \Phi_B \, , \tag{10}$$

The details of this computation are presented in Ref. 22.

## III. RESULTS AND DISCUSSIONS

Electronic properties of CNRs along various orientations have been widely studied[7,8,23,24]. Using simple $\pi$-orbital tight-binding approaches, zigzag CNRs and armchair CNRs have been predicted to have metallic and semiconducting properties, respectively. Although a recent theoretical study showed that zigzag CNRs can also have a small bandgap when spin effects are considered[25], the bandgap is too small to be used for MOSFET-type devices. In this work, therefore, we focus on exploring the physical properties and device performance of armchair CNR MOSFETs.

### a) 1D perfect contacts vs. 2D semi-infinite contacts

The nature of the contacts plays an important role in the transport properties of nanostructures. We explore the dimensional effect of the contacts on the properties of the armchair CNR devices shown in Fig. 2. Figures 2(a), (b), and (c) present the device structure, transmission vs. energy, $T(E)$ and density of states vs. energy, $DOS(E),$ for 12.5 nm long and 1.4nm wide armchair CNR with 1D contacts. Due to the perfect contacts, the device behaves just like a homogenous structure in equilibrium. The staircase in $T(E)$ and sharp peak in $DOS(E)$ demonstrate the infinite 1D material's characteristics. As the CNR becomes wider, the shapes of $T(E)$ and $DOS(E)$ will become closer to that of 2D graphene sheets, cf., Fig. 1, because the quantum confinement effects become weaker and eventually lose their importance in large size device structures.

Figures 2(d), (e), and (f) show the device structure, transmission vs. energy, *T(E)* and density of states vs. energy, *DOS(E),* respectively, for 60 nm long and 1.4 nm wide armchair CNR with *2D* contacts. Compared to the results of the CNR device with 1D contacts, the overall shapes of *T(E)* and *DOS(E)* of two cases are similar. The small oscillations on the curves are attributed to reflections from the interfaces between the ribbon and the graphene half-plane. A localized peak appears in middle of band gap in Fig. 2(f) resulting from the metallic property of the 2D graphene sheet. Since the Fermi level of the intrinsic CNR also appears at midgap, the metal-induced gap states are expected to cause Fermi level pinning and degrade the device performance. Although these do not affect the transmission (Fig. 2(e)) for long channel devices under equilibrium, they are expected to contribute to tunneling currents in the short channel devices or even in the long channel devices due to scattering processes under drain bias. In this work, in order to evaluate the ultimate performance of a ballistic CNR MOSFETs, we would focus on the armchair CNR MOSFETs with 1D perfect contacts.

b) Performance of armchair CNR MOSFETs

To explore the performance of armchair CNR MOSFETs with 1D contacts, a double gate MOSFET structure as shown in Fig. 3 is used. Figure 3(a) shows the side view of the device, where the CNR is placed within two insulator layers, assumed to be $SiO_2$ of 1 nm thickness. Gate electrodes are placed in the top and bottom of the insulators. The source and drain regions are assumed to be doped CNR regions with 0.08 electrons per carbon atom, corresponding to $1.37 \times 10^8$ electrons/m. Figure 3(b) shows the top view of the CNR channel. In this work, a 1.4 nm wide armchair CNR, whose bandgap is around 0.8 eV, is used to explore the ON-currents and OFF-currents performance of the MOSFET-type device. The channel is assumed to be 12.5 nm long and undoped.

Using the NEGF approach, the local density of states vs. energy and position, *LDOS(x,E)*, is calculated. We computed the density of states of the device according to the realistic atomic positions, but plot *LDOS(x,E)* by averaging the value per atom in an unit cell. Fig. 4(a) shows the *LDOS(x,E)* in the device under equilibrium, including parts of the source and drain. The upper and lower dashed lines correspond to the energies of the edges of the first conduction subband and the first valence subband of the device. In the conduction band region, the oscillation patterns are attributed to quantum mechanical reflections, whereas the second and third subbands are clearly visible. The separations of the subbands are around 0.5 and 0.7 eV which agrees with the subband separations obtained from the dispersion relations for electrons in an infinitely long 1.4 nm wide armchair CNR. The strong oscillation patterns of the LDOS in the each subband visible in the source and drain are similar to those found in previous theoretical studies of a semiconductor nano-MOSFETs and CNT MOSFETs[15,26,27]. They occur due to the quantum reflection off the barrier of the channel. The light-grey colored area in the bandgap under the first conduction band, especially inside the channel region, corresponds to the density of states caused by the evanescent tail of the electron wave function inside the source and drain that penetrates into the undoped channel region. Under certain operating bias, these states could contribute to tunneling and degrade the device performance. In the valence band region, the localized states can be observed clearly due to quantum confinement effects cased by the quantum well. These states would also contribute to tunneling currents when the top of the valence band is close to the bottom of the conduction band of the source, for example under strong negative gate biases. They could be washed-out by increasing the channel length or CNR width. These states play an important role to band-to-band tunneling devices.

Next, we plot the current density in the transport direction of the channel vs. energy, *J(x,E)*. Similarly to *LDOS(x,E)*, the averaged *J(x,E)* in a unit cell under $V_{DS}$=0.4V and $V_{GS}$=0.7 V, normalized by $q^2/h$, is shown in Fig. 4(b). Due to the ballistic transport assumed, the current density is constant throughout the entire length of the device. The main contribution to the current comes from the energy window region between the chemical potential of the source and the top of the barrier between the source and channel. Quantum simulations capture both thermionic emitted and tunneling currents which can be important. More details about tunneling current will be discussed later.

The source-to-drain current vs. $V_{DS}$ for different $V_{GS}$ values of the armchair CNR MOSFET is shown in Fig. 5(a). The results show that CNR MOSFET has good MOSFET-type device behavior. The current densities of these CNR MOSFETs at $V_{GS}$=0.6 V and $V_{GS}$=0.7 V are around 2500 and 4200 μA/μm respectively, satisfying the requirement of the 2006 ITRS report for the year 2015[28]. Note that our simulations use thicker insulators and smaller power supply (0.4 V) compared to the projections of the 2006 ITRS report. Figure 5(b) shows the $I_{DS}$-$V_{GS}$ characteristics of a 1.4 nm and 1.8 nm wide armchair CNR MOSFETs with a channel length of 12.5 nm. The corresponding bandgaps are $E_G$=0.8 eV and $E_G$=0.66 eV respectively. The DIBL in these two devices is 30 mV/V and 60mV/V respectively, whereas the subthreshold swings are 68mV/dec and 74mV/decad. These values are smaller than DIBL=122 mV/V and S=90 mV/decade, that are the estimated values of a double gate, 10-nm-scaled Si MOSFETs[14]. This can be attributed to the better gate control on the electrostatics of the CNR MOSFET device compared to Si MOSFETs. CNR is a monolayer material, i.e. the ultimate ultra-thin body channel. Therefore, 2D-elecrostatic effects could be suppressed and DIBL can be reduced. However, it still cannot reach the fundamental thermal limit, i.e. $S$=60mV/decade at room temperature, because the light effective mass of carriers in CNR, and the short channel length in combination with the small bandgap, enhance source-to-drain tunneling. As a result, the OFF state current of the 1.8 nm wide MOSFET is around two orders of magnitude higher than that of the 1.4 nm wide MOSFET, and its' $S$ higher too. This quantum tunneling current plays a significantly important role in the OFF state of MOSFET and $SS$ degradation.

Fig. 6 shows the ratio of quantum tunneling current (solid line) and thermionic (dashed line) current to the total current vs. $V_{GS}$ bias. The tunneling current, is calculated from current contribution under the top of the barrier, and the thermionic current is calculated from the current contribution above the top of the barrier. Both results are presented at $V_{DS}$=0.4V (diamonds) and $V_{DS}$=0.05V (circles) for $V_{GS}$ from -0.6 V to 0.7 V. We found that, when $V_{GS}$ decreases (more negative bias), the ratio of quantum tunneling current to the total current increases, whereas the ratio of thermally emitted current to the total current decreases. It is because the barrier height increases as $V_{GS}$ decreases, reducing all thermally emitted contribution. When $V_{GS}$ is less than -0.2V, the tunneling current due to band-to-band tunneling mechanisms completely dominates, and thus determines the OFF-current. As $V_{GS}$ increases (more positive bias), the tunneling current decreases (since the band to band tunneling mechanisms are inhibited), but around 0.3V it increases again due to the quantum tunneling currents under the top of the barrier. Thus the tunneling current plays a less important role when the device is in the ON-state (under high gate bias) because most of electrons would go above the top of the barrier, just like in a classical MOSFET.

The tunneling current, however, still cannot be ignored because it contributes around 18% of the total current even at ON-state conditions ($V_G$=0.7V and $V_{DS}$=0.4V).

Figure 7 shows the electron density vs. energy along the length of the device, $G^n(x,E)$ at $V_D$=0.4V and $V_G$=-0.6V. The quantum states in the valence band provide the paths to achieve band-to-band (BTBT) currents. Once these valence band states rise entirely above the bottom of the conduction band of the source (stronger negative gate bias), the current will increase dramatically. This phenomenon accounts for the rise in current at negative bias, see e.g. below $V_G$=-0.2V in Fig. 7. This behavior is similar to that observed[14] and simulated[15] for carbon nanotube transistors. The device could then be designed for BTBT MOSFET operation. Since the thermionic-tunneling current can be ignorable in these operating conditions, the *SS* would not follow the thermal limit. Therefore, BTBT MOSFETs are expected to outperform the normal MOSFETs in terms of *SS* behavior (sub 60mV/dec at room temperature), something experimentally demonstrated for CNT MOSFETs. (In our simulations there is a small indication of this sub-60mV/dec behavior which can be further enhanced, but an analysis and optimization of a band-to-band tunneling type of device[29] is beyond the scope of this paper). Due to similar electronic structures between CNRs and CNTs, CNRs would also have the potential for BTBT-type of MOSFETs applications. This type of device however, might suffer from the low drive current capabilities due to the reduced magnitude of the tunneling currents.

IV. CONCLUSION

In this work, we developed the real space quantum transport simulation for CNR FETs using NEGF approach based on a $\pi$-orbital TB method. The model is used to investigate the interface properties of an armchair CNR with 1D and 2D contacts. The localized states caused by the semi-metallic 2D graphene sheet are observed. Since these states appear in the middle of the bandgap, they would cause Fermi level pinning and degrade the device performance. Moreover, we investigate the device performance of a ballistic armchair CNR MOSFETs with 1.4 nm and 1.8 nm width, both of 12.5nm channel length. The device structure is a double gate MOSFET structure configuration of 1nm insulator thickness. We found that these CNR MOSFETs can have better performance than a double gate, 10-nm-scale Si MOSFETs in terms of $S$ and DIBL. Although tunneling processes cannot be ignorable and degrade the device performance for usual CNR MOSFET-type device, it can have potential applications to CNR MOSFETs operating in the BTBT MOSFET mode.

ACKNOWLEDGEMENTS

The work at national university of Singapore was supported by MOM under grant R-263-000-416-112 and R-263-000-416-133. The work at Purdue University was supported by the Semiconductor Research Corporation (SRC). The authors would like to thank Siyuranga Koswatta for helpful discussions and the national Science Foundation's Network for Computational Nanotechnology (NCN) for providing computational resources.

FIGURE CAPTIONS

Fig.1: Density of states and transmission of an infinite 2D Graphite sheet vs. Energy/$\tau$, where $\tau$ is the $\pi$ orbital coupling of a tight-binding model. (Obtained by using the recursive surface Green's function approach). The results are in a good agreement with Ref. 16.

Fig. 2: (a) and (d) Schematics of the top view of the device structures for 1D and 2D contacts, respectively. (b,c) Transmission coefficient and density of states, respectively, vs. energy of a 1.4 nm wide CNR with 1D contacts showing the perfect 1D transmission and DOS. (e,f) Transmission coefficient and density of states, respectively, vs. energy of a 1.4 nm CNR with 2D infinite contacts (semi-infinite graphene sheet).

Fig. 3: (a) A schematic diagram of the simulated dual-gate carbon nanoribbon MOSFETs. The source and drain are heavily doped nanoribbon contacts while the channel is undoped. The oxide thickness ($t_{ins}$) is 1 nm in this study. (b) Top view of (a).

Fig. 4: (a) The local density of states, LDOS(x,E), at equilibrium. Dashed lines show the band profile of the lowest conduction band and the highest valence band of 1.4nm width armchair CNR. The second and third conduction subbands are Cleary seen in the plot, as well as quantum levels in the valence band due to longitudinal confinement. (b) The ballistic current density, $J(x,E)$, normalized by $q^2/h$, for $V_G$=0.7 V and $V_D$=0.4V. Current contribution comes from electrons above the top of the barrier and below the Fermi level of the source ($E_{FS}$=0 in this case).

Fig. 5: (a) Simulated current $I_{DS}$ vs. $V_{DS}$ for a 1.4 nm wide and 12.5 nm channel long armchair CNR at $V_{GS}$=0.4, 0.5, 0.6 and 0.7 V. (b) Simulated current $I_{DS}$ vs. $V_{GS}$ for a 1.4 nm wide and 12.5 nm long channel armchair CNR at $V_{DS}$=0.05 and 0.4V, as well as for a 1.8 nm wide and 12.5 nm long armchair CNR at $V_{DS}$=0.4V.

Fig. 6: Ratio of the tunneling current (solid line) and thermionic current (dashed line) to the total current at $V_{DS}$=0.05 V (circle) and $V_{DS}$= 0.4 V (diamond) for $V_{GS}$ from -0.6 V to 0.7 V. The quantum tunneling current plays an important role in the OFF-current when the gate voltage decreases. When $V_{GS}$ is smaller than -0.3 V, the tunneling current dominates the OFF current.

Fig. 7: Simulated $N(x,E)$ at $V_{DS}$=0.4 V and $V_{GS}$=-0.6V. When the quantum levels in the channel's valence band are close to the bottom of source's conduction band, more electrons start tunneling between bands and current starts to increase.

Figure 1

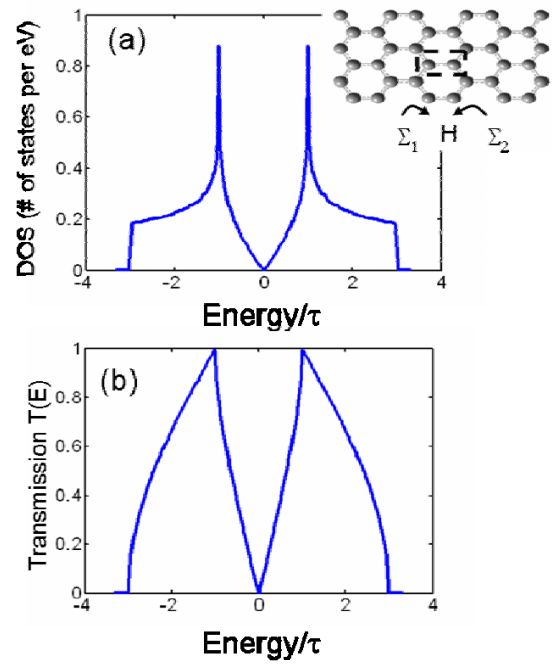

Figure 2

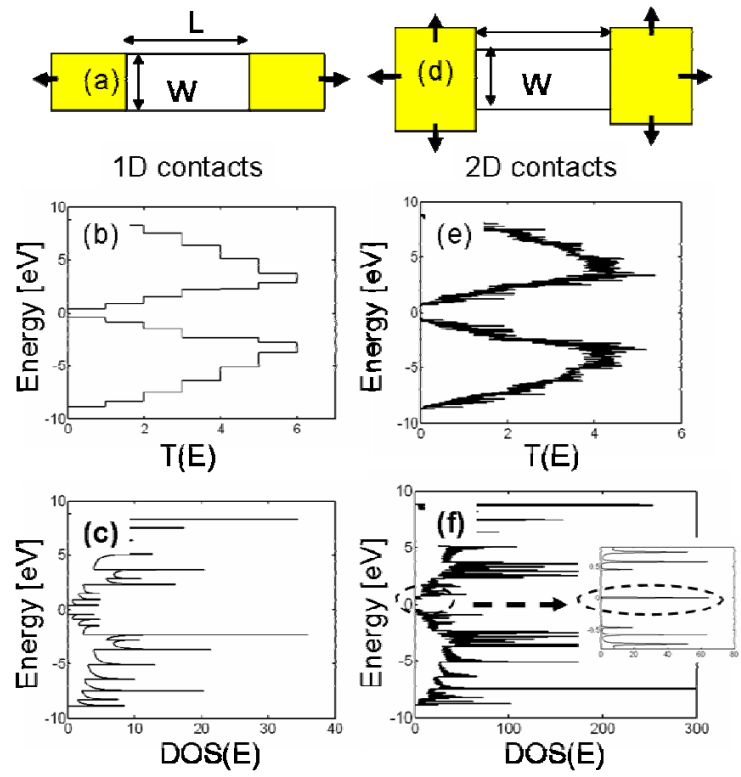

Figure 3

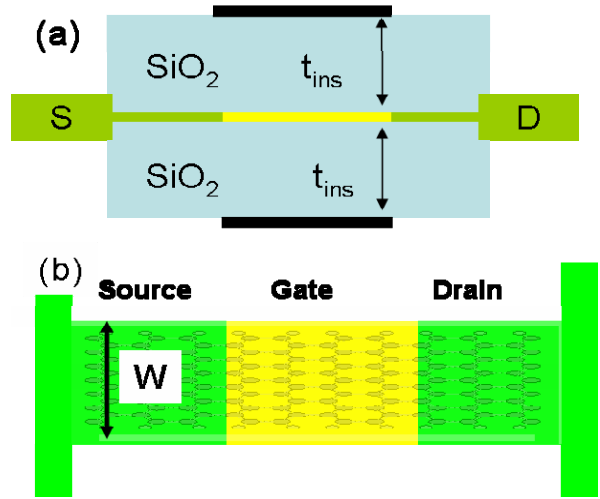

Figure 4

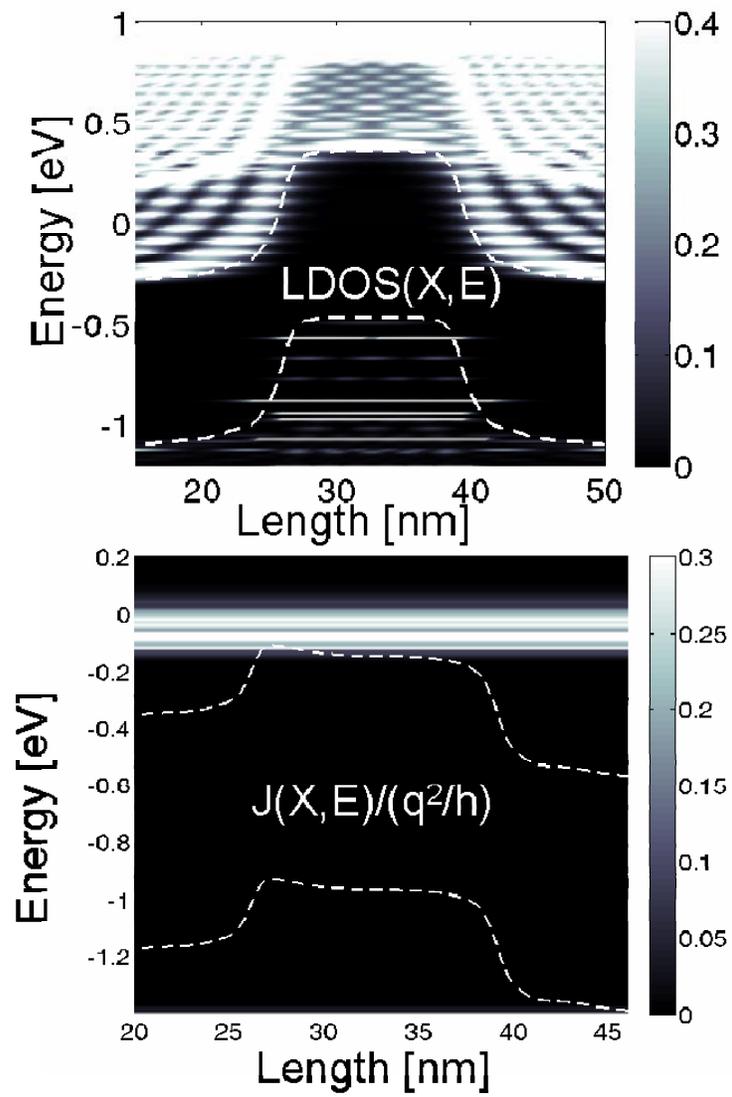

Figure 5

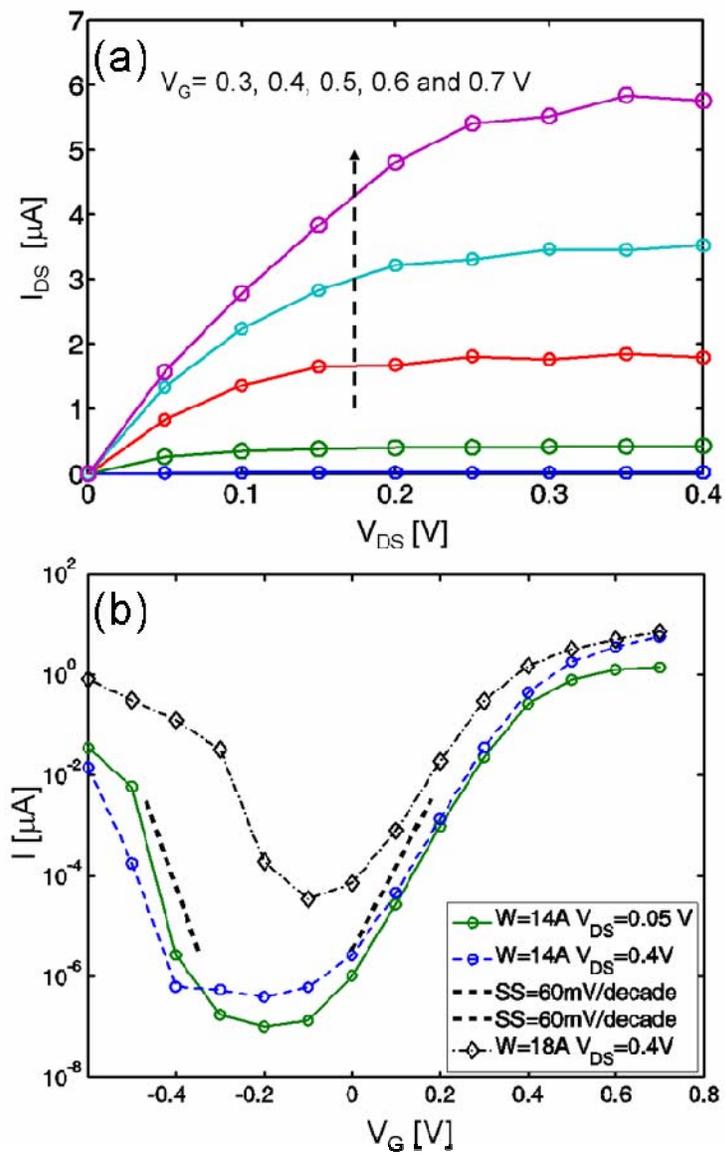

Figure 6

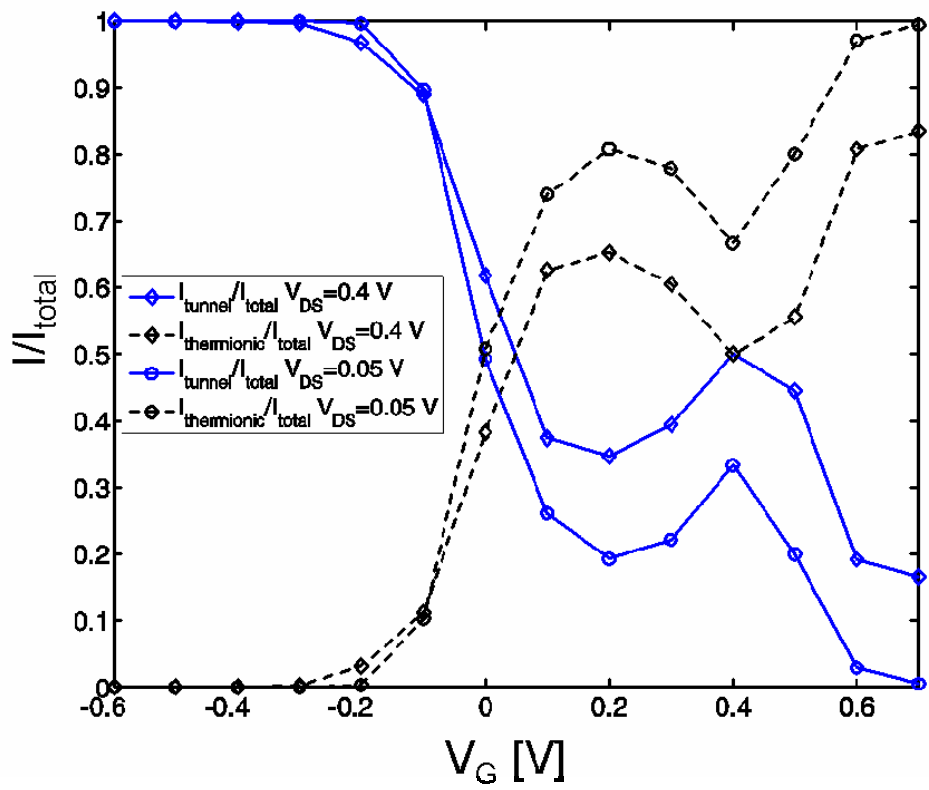

Figure 7

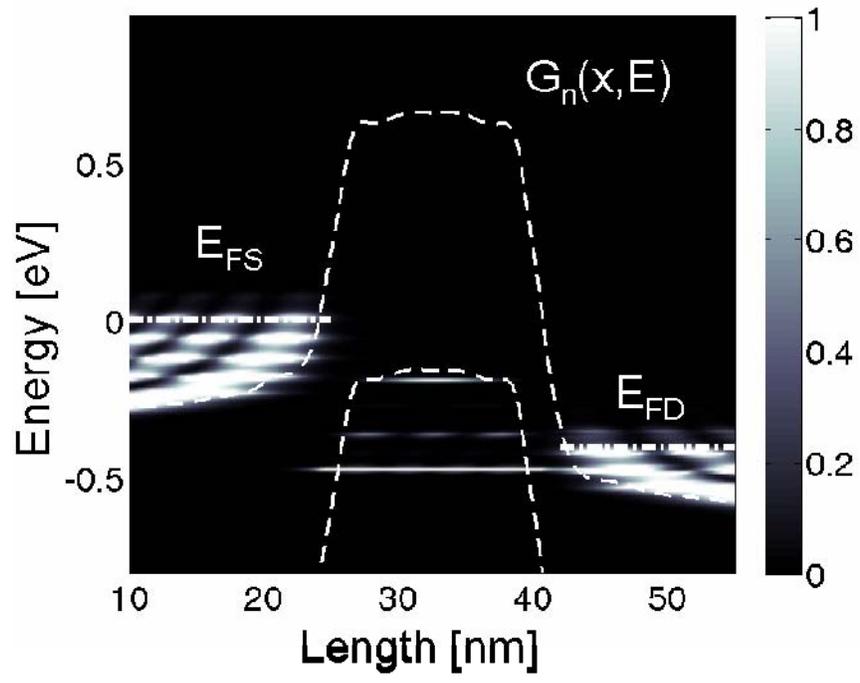